# Comment on "Conjugate heat transfer of mixed convection for viscoelastic fluid past a stretching sheet", by Kai-Long Hsiao and Guan-Bang Chen [Mathematical Problems in Engineering, Volume 2007, article 17058, 21 pages]

Asterios Pantokratoras

School of Engineering, Democritus University of Thrace,
67100 Xanthi – Greece
e-mail:apantokr@civil.duth.gr

In the above paper an analysis has been carried for the flow along a vertical linearly stretching sheet taking into account the heat conduction of the sheet. The sheet temperature varies in a linear way whereas the fluid is viscoelastic. The boundary layer equations are transformed into ordinary ones and subsequently are solved numerically. However, there are some errors in the above paper which are presented below:

1. The basic parameters used in the above paper are the buoyancy parameter G, the viscoelastic parameter E, the Pr number and the conduction-convection parameter $N_\infty$. In the energy equation 2.6 a source term is included ($q(T-T_\infty)$). However, this source term has been ignored because no non-dimensional parameter exist in the paper which expresses the source term of the energy equation.
2. The quantities $f^{IV}$ (equation 2.9), $Re_x$ (equation 2.20) and $T_e$ (equation 2.25) have not been defined in the paper.
3. In the denominator of equations (2.20) and (2.22) a velocity $u_\infty$ appears. Usually, in boundary layer theory, as $u_\infty$ is defined the fluid velocity of the ambient fluid. However, in the present problem the ambient fluid velocity is zero (equation 2.7) and therefore the quantities in equations (2.20) and (2.22) can not be defined with a zero denominator.
4. The caption of figure 4.3 contains the sentence "Dimensionless temperature profiles $\theta'(0)$ versus $\eta$ as $G = 1.0$, $E = 0.001$ and Pr =0.001,0.7,2.0,10.0". However the quantity $\theta'(0)$ is the temperature gradient at the plate and does not change along $\eta$.

5. The caption of figure 4.4 contains the sentence "Dimensionless temperature profiles θ(0) versus η as $G = 1$, $E = 0.0001$ and Pr $=0.001, 10$". However the quantity θ(0) is the temperature at the plate and does not change along η.
6. The caption of figure 4.5 contains the sentence "Dimensionless temperature gradient profiles θ'(0) versus η as $G = 1$, $E = 0.001$ and Pr $=0.001, 0.7, 2, 10$". However the quantity θ'(0) is the temperature gradient at the plate and does not change along η.
7. The caption of figure 4.6 contains the sentence "Dimensionless temperature profiles θ(0) versus η as $G = 1$, $E = 0.001$ and Pr $=0.001, 10$". However the quantity θ(0) is the temperature at the plate and does not change along η.
8. The caption of figure 4.7 contains the sentence "Dimensionless temperature gradient profiles θ'(0) versus η as $G = 1$, $E = 0.01$ and Pr $=0.001, 0.7, 2, 10$". However the quantity θ'(0) is the temperature gradient at the plate and does not change along η.
9. The caption of figure 4.8 contains the sentence "Dimensionless temperature profiles θ(0) versus η as $G = 1$, $E = 0.01$ and Pr $=0.001, 0.7, 2, 10$". However the quantity θ(0) is the temperature at the plate and does not change along η.
10. The caption of figure 4.9 contains the sentence "Dimensionless temperature gradient profiles θ'(0) versus η as $G = 1$-$25$, $E = 0.001$-$0.15$ and Pr $=1$". However the horizontal axis contains the Pr number and not the the quantity η.
11. In figures 4.4, 4.6 and 4.8 the wide temperature profiles (those that do not approach the horizontal axis asymptotically) are truncated due to a small calculation domain used (see Pantokratoras, 2008b). Although the authors recognize the problem with these profiles (page 11) they did not solve it and these three profiles are wrong.
12. It is known in boundary layer theory that as the Pr number increases the temperature profiles become thinner and vice versa when the Pr number decreases the temperature profiles become thicker. See Arpaci and Larsen (1984, page 225), Chen and Char (1988, page 575), Cortell (2007, page 870), Cortell (2008, page 1342), Kakac and Yener (1995, page 325), Pantokratoras (2004, page 1895), Pantokratoras (2008a, page 108), Schlichting and Gersten (2003, page 215), Siddheshwar and Mahabaleswar (2005, page 818),

Subhas Abel et al. (2007, page 964), Tsou et al. (1967, page 222), Wang (1994, page 60) and White (2006, page 325). In the above paper this principle has been violated. In figures 4.4, 4.6 and 4.8 the temperature profiles become thinner as the Pr number decreases and thicker as the Pr number increases. This is a serious error.